\documentclass[twocolumn,preprintnumbers,nofootinbib,prd]{revtex4}
\usepackage{psfrag,graphicx,epsfig,amsmath,amssymb,bm}
\begin{document}

\preprint{MZ-TH/09-25} 

\title{Two-loop divergences of scattering amplitudes with massive partons}

\author{Andrea Ferroglia, Matthias Neubert, Ben D.~Pecjak, and Li Lin Yang} 

\affiliation{Institut f\"ur Physik (THEP), Johannes Gutenberg-Universit\"at, D-55099 Mainz, Germany}
\date{\today}

\begin{abstract}
\noindent
We complete the study of two-loop infrared singularities of scattering amplitudes with an arbitrary number of massive and massless partons in non-abelian gauge theories. To this end, we calculate the universal functions $F_1$ and $f_2$, which completely specify the structure of three-parton correlations in the soft anomalous-dimension matrix, at two-loop order in closed analytic form. Both functions are found to be suppressed like ${\cal O}(m^4/s^2)$ in the limit of small parton masses, in accordance with mass factorization theorems proposed in the literature. On the other hand, they are unsuppressed and diverge logarithmically near the threshold for pair production of two heavy particles. As an application, we calculate the two-loop anomalous-dimension matrix for $q\bar q\to t\bar t$ near threshold and show that it is not diagonal in the $s$-channel singlet-octet basis.
\end{abstract}

\maketitle

\section{Introduction}

Recently, much progress has been achieved in the understanding of the infrared (IR) singularities of massless scattering amplitudes in non-abelian gauge theories. While factorization proofs guarantee the absence of IR divergences in inclusive observables \cite{Collins:1989gx}, in many cases large Sudakov logarithms remain after this cancellation. A detailed control over the structure of IR poles in the virtual corrections to scattering amplitudes is a prerequisite for the resummation of these logarithms beyond the leading order \cite{Sterman:1986aj,Catani:1989ne,Contopanagos:1996nh,Kidonakis:1997gm}. Catani was the first to  predict the singularities of two-loop scattering amplitudes apart from the $1/\epsilon$ pole term \cite{Catani:1998bh}, whose general form was only understood much later in \cite{Sterman:2002qn,MertAybat:2006wq,MertAybat:2006mz}. In recent work \cite{Becher:2009cu}, it was shown that the IR singularities of on-shell amplitudes in massless QCD can be derived from the ultraviolet (UV) poles of operator matrix elements in soft-collinear effective theory (SCET). They can be subtracted by means of a multiplicative renormalization factor, whose structure is constrained by the renormalization group. It was proposed in this paper that the simplicity of the corresponding anomalous-dimension matrix holds not only at one- and two-loop order, but may in fact be an exact result of perturbation theory. This possibility was raised independently in \cite{Gardi:2009qi}. Detailed theoretical arguments supporting this conjecture were presented in \cite{Becher:2009qa}, where constraints derived from soft-collinear factorization, the non-abelian exponentiation theorem, and the behavior of scattering amplitudes in two-parton collinear limits were studied.

It is relevant for many physical applications to generalize these results to the case of massive partons. The IR singularities of one-loop amplitudes containing massive partons were obtained some time ago in \cite{Catani:2000ef}, but until very recently little was known about higher-loop results. In the limit where the parton masses are small compared with the typical momentum transfer among the partons, mass logarithms can be predicted based on collinear factorization theorems \cite{Mitov:2006xs,Becher:2007cu}. This allows one to obtain massive amplitudes from massless ones with a minimal amount of calculational effort. A major step toward solving the problem of finding the IR divergences of generic two-loop scattering processes with both massive and massless partons has been taken in \cite{Mitov:2009sv,Becher:2009kw}. One finds that the simplicity of the anomalous-dimension matrix observed in the massless case no longer persists in the presence of  massive partons. Important constraints from soft-collinear factorization and two-parton collinear limits are lost, and only the non-abelian exponentiation theorem restricts the allowed color structures in the anomalous-dimension matrix. At two-loop order, two different types of three-parton color and momentum correlations appear, whose effects can be parameterized in terms of two universal, process-independent functions $F_1$ and $f_2$ \cite{Becher:2009kw}. Apart from some symmetry properties, the precise form of these functions was left unspecified. In this Letter we calculate these functions at two-loop order and study their properties in some detail.

\section{Anomalous-dimension matrix}

We denote by $|{\cal M}_n(\epsilon,\{\underline{p}\},\{\underline{m}\})\rangle$ a UV-renormalized, on-shell $n$-parton scattering amplitude with IR singularities regularized in $d=4-2\epsilon$ dimensions. Here $\{\underline{p}\}\equiv\{p_1,\dots,p_n\}$ and $\{\underline{m}\}\equiv\{m_1,\dots,m_n\}$ denote the momenta and masses of the external partons. The amplitude is a function of the Lorentz invariants $s_{ij}\equiv 2\sigma_{ij}\,p_i\cdot p_j+i0$ and $p_i^2=m_i^2$, where the sign factor $\sigma_{ij}=+1$ if the momenta $p_i$ and $p_j$ are both incoming or outgoing, and $\sigma_{ij}=-1$ otherwise. For massive partons we define 4-velocities $v_i=p_i/m_i$ with $v_i^2=1$ and $v_i^0\ge 1$. We further define the recoil variables $w_{ij}\equiv-\sigma_{ij}\,v_i\cdot v_j-i0$. We use the color-space formalism \cite{Catani:1996jh}, in which $n$-particle amplitudes are treated as $n$-dimensional vectors in color space. $\bm{T}_i$ is the color generator associated with the $i$-th parton and acts as a matrix on its color index. The product $\bm{T}_i\cdot\bm{T}_j\equiv T_i^a\,T_j^a$ is summed over $a$. Generators associated with different particles commute, and $\bm{T}_i^2=C_i$ is given in terms of the eigenvalue of the quadratic Casimir operator of the corresponding color representation, i.e., $C_q=C_{\bar q}=C_F$ for quarks and $C_g=C_A$ for gluons. Below, we will label massive partons with capital indices ($I,J,\dots$) and  massless ones with lower-case indices ($i,j,\dots$). 

It was shown in \cite{Becher:2009cu,Becher:2009qa,Becher:2009kw} that the IR poles of such amplitudes can be removed by a multiplicative renormalization factor $\bm{Z}^{-1}(\epsilon,\{\underline{p}\},\{\underline{m}\},\mu)$, which acts as a matrix on the color indices of the partons. More precisely, the product $\bm{Z}^{-1}|{\cal M}_n\rangle$ is finite for $\epsilon\to 0$ after the coupling constant $\alpha_s^{\rm QCD}$ used in the calculation of the scattering amplitude is properly matched onto the coupling $\alpha_s$ in the effective theory, in which the heavy partons are integrated out \cite{Becher:2009kw}. The relation
\begin{equation}\label{RGE}
   \bm{Z}^{-1}\,\frac{d}{d\ln\mu}\,
   \bm{Z}(\epsilon,\{\underline{p}\},\{\underline{m}\},\mu) 
   = - \bm{\Gamma}(\{\underline{p}\},\{\underline{m}\},\mu)
\end{equation}
links the renormalization factor to a universal anomalous-dimension matrix $\bm{\Gamma}$, which governs the scale dependence of effective-theory operators built out of collinear SCET fields for the massless partons and soft heavy-quark effective theory fields for the massive ones. For the case of massless partons, the anomalous dimension has been calculated at two-loop order in \cite{MertAybat:2006wq,MertAybat:2006mz} and was found to contain only two-parton color-dipole correlations. It has recently been conjectured that this result may hold to all orders of perturbation theory \cite{Becher:2009cu,Gardi:2009qi,Becher:2009qa}. On the other hand, when massive partons are involved in the scattering process, then starting at two-loop order correlations involving more than two partons appear \cite{Mitov:2009sv}. At two-loop order, the general structure of the anomalous-dimension matrix is \cite{Becher:2009kw}
\begin{eqnarray}\label{resu1}
   \bm{\Gamma}
   &=& \sum_{(i,j)}\,\frac{\bm{T}_i\cdot\bm{T}_j}{2}\,
    \gamma_{\rm cusp}(\alpha_s)\,\ln\frac{\mu^2}{-s_{ij}}
    + \sum_i\,\gamma^i(\alpha_s) \nonumber\\
   &&\mbox{}- \sum_{(I,J)}\,\frac{\bm{T}_I\cdot\bm{T}_J}{2}\,
    \gamma_{\rm cusp}(\beta_{IJ},\alpha_s)
    + \sum_I\,\gamma^I(\alpha_s) \nonumber\\
   &&\mbox{}+ \sum_{I,j}\,\bm{T}_I\cdot\bm{T}_j\,
    \gamma_{\rm cusp}(\alpha_s)\,\ln\frac{m_I\mu}{-s_{Ij}} \\
   &&\mbox{}+ \sum_{(I,J,K)} if^{abc}\, 
    \bm{T}_I^a\,\bm{T}_J^b\,\bm{T}_K^c\,
    F_1(\beta_{IJ},\beta_{JK},\beta_{KI}) \nonumber\\
   &&\mbox{}+ \sum_{(I,J)} \sum_k\,if^{abc}\,
    \bm{T}_I^a\,\bm{T}_J^b\,\bm{T}_k^c\,
    f_2\Big(\beta_{IJ},
    \ln\frac{-\sigma_{Jk}\,v_J\cdot p_k}{-\sigma_{Ik}\,v_I\cdot p_k}
    \Big) \,. \nonumber
\end{eqnarray}
The one- and two-parton terms depicted in the first three lines start at one-loop order, while the three-parton terms in the last two lines appear at ${\cal O}(\alpha_s^2)$. The notation $(i,j,\dots)$ etc.\ means that the corresponding sum extends over tuples of distinct parton indices. The cusp angles $\beta_{IJ}$ are defined via
\begin{equation}\label{wbrela}
   \cosh\beta_{IJ} = \frac{-s_{IJ}}{2m_I m_J} = w_{IJ} \,.
\end{equation}
They are associated with the hyperbolic angles formed by the time-like Wilson lines of two heavy partons. The physically allowed values are $w_{IJ}\ge 1$ (one parton incoming and one outgoing), corresponding to $\beta_{IJ}\ge 0$, or $w_{IJ}\le -1$ (both partons incoming or outgoing), corresponding to $\beta_{IJ}=i\pi-b$ with real $b\ge 0$. These  possibilities correspond to space-like and time-like kinematics, respectively. Since in a three-parton configuration there is always at least one pair of partons either incoming or outgoing, at least one of the $w_{IJ}$ or $v_I\cdot p_k$ variables must be time-like, and hence the functions $F_1$ and $f_2$ have non-zero imaginary parts. 

The anomalous-dimension coefficients $\gamma_{\rm cusp}(\alpha_s)$ and $\gamma^i(\alpha_s)$ (for $i=q,g$) in (\ref{resu1}) have been determined to three-loop order in \cite{Becher:2009qa} by considering the case of the massless quark and gluon form factors. Similarly, the coefficients $\gamma^I(\alpha_s)$ for massive quarks and color-octet partons such as gluinos have been extracted at two-loop order in \cite{Becher:2009kw} by analyzing the anomalous dimension of heavy-light currents in SCET. In addition, the velocity-dependent function $\gamma_{\rm cusp}(\beta,\alpha_s)$ has been derived from the known two-loop anomalous dimension of a current composed of two heavy quarks moving at different velocity \cite{Korchemsky:1987wg,Kidonakis:2009ev}.

Here we complete the calculation of the two-loop anomalous-dimension matrix by deriving closed analytic expressions for the universal functions $F_1$ and $f_2$, which parameterize the three-parton correlations in (\ref{resu1}).

\section{Calculation of $\bm{F_1}$ and $\bm{f_2}$}
\label{sec:F1f2}

To calculate the function $F_1$ we compute the two-loop vacuum matrix element of the operator $\bm{O}_s=\bm{S}_{v_1}\,\bm{S}_{v_2}\,\bm{S}_{v_3}$, which consists of three soft Wilson lines along the directions of the velocities of three massive partons, without imposing color conservation. The anomalous dimension of this operator contains a three-parton term given by $6if^{abc}\,\bm{T}_1^a\,\bm{T}_2^b\,\bm{T}_3^c\,F_1(\beta_{12},\beta_{23},\beta_{31})$. The function $F_1$ follows from the coefficient of the $1/\epsilon$ pole in the bare matrix element of $\bm{O}_s$. We will then obtain $f_2$ from a limiting procedure.

\begin{figure}
\begin{center}
\includegraphics[width=0.32\columnwidth]{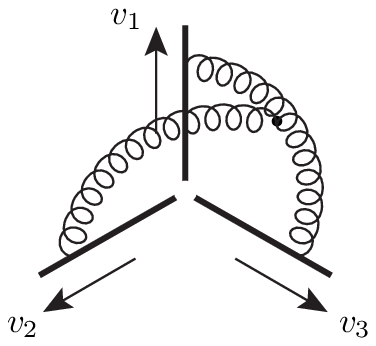}
\includegraphics[width=0.63\columnwidth]{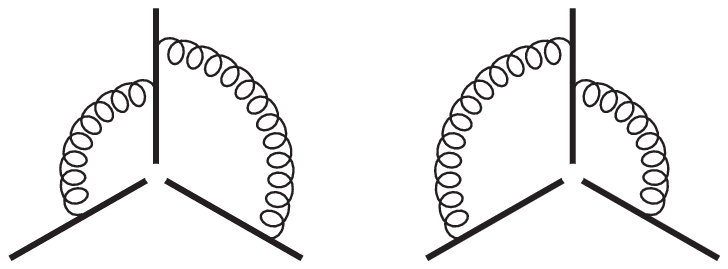}
\includegraphics[width=0.63\columnwidth]{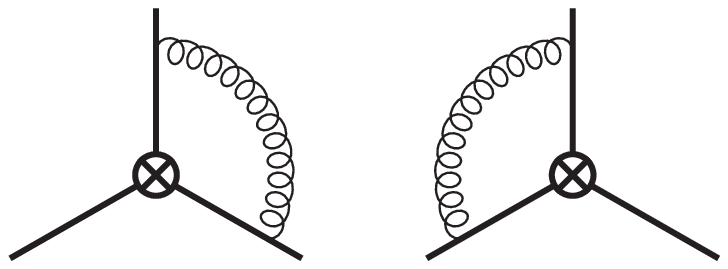}
\caption{\label{fig:dia}
Two-loop Feynman graphs (top row) and one-loop counterterm diagrams (bottom row) contributing to the two-loop coefficient of the renormalization factor $\bm{Z}_s$.}
\end{center}
\vspace{-4mm}
\end{figure}

The operator $\bm{O}_s$ is renormalized multiplicatively, so that $\bm{O}_s\bm{Z}_s$ is UV finite, where $\bm{Z}_s$ is linked to the anomalous dimension in the same way as shown in (\ref{RGE}). In order to calculate the two-loop $\bm{Z}_s$ factor, we have evaluated the two-loop non-planar and planar graphs shown in the first row of Figure~\ref{fig:dia}, as well as the one-loop counterterm diagrams depicted in the second row. Contrary to a statement made in \cite{Mitov:2009sv}, we find that $F_1$ receives contributions from all five diagrams, not just from the non-planar graph. The most challenging technical aspect of the analysis is the calculation of the diagram involving the triple-gluon vertex. We have computed this diagram using a Mellin-Barnes representation and checked the answer numerically using sector decomposition \cite{Smirnov:2008py}. We have also checked that for Euclidean velocities our result for the triple-gluon diagram agrees numerically with a position-space based integral representation derived in \cite{Mitov:2009sv}. Combining all contributions, we find
\begin{equation}\label{eq:F1}
   F_1(\beta_{12},\beta_{23},\beta_{31}) 
   = \frac{\alpha_s^2}{12\pi^2}
    \sum_{i,j,k} \epsilon_{ijk}\,g(\beta_{ij})\,r(\beta_{ki}) \,,
\end{equation}
where we have introduced the functions
\begin{eqnarray}
   r(\beta) 
   &=& \beta\,\coth\beta \,, \nonumber\\
   g(\beta) 
   &=& \coth\beta \left[ \beta^2 
    + 2\beta\,\ln(1-e^{-2\beta}) - \mbox{Li}_2(e^{-2\beta}) 
    + \frac{\pi^2}{6} \right] \nonumber\\
   &&\mbox{}- \beta^2 - \frac{\pi^2}{6} \,.
\end{eqnarray}
The function $f_2$ can be obtained from the above result by writing $w_{23}=-\sigma_{23}\,v_2\cdot p_3/m_3$, $w_{31}=-\sigma_{31}\,v_1\cdot p_3/m_3$ and taking the limit $m_3\to 0$ at fixed $v_I\cdot p_3$. In that way, we obtain
\begin{equation}
   f_2\Big( \beta_{12}, 
    \ln\frac{-\sigma_{23}\,v_2\cdot p_3}%
            {-\sigma_{13}\,v_1\cdot p_3} \Big) 
   = - \frac{\alpha_s^2}{4\pi^2}\,g(\beta_{12})\,
    \ln\frac{-\sigma_{23}\,v_2\cdot p_3}%
            {-\sigma_{13}\,v_1\cdot p_3} \,.
\end{equation}
Whether a factorization of the three-parton terms into two functions depending on only a single cusp angle persists at higher orders in $\alpha_s$ is an open question. 

It is interesting to expand the two functions $r(\beta)$ and $g(\beta)$ about the threshold point $\beta=i\pi-b$ with $b\to 0^+$. We find
\begin{equation}\label{rgthreshold}
\begin{aligned}
   r(\beta)
   &= - \frac{i\pi}{b} + 1 + {\cal O}(b) \,, \\
   g(\beta)
   &= - \frac{\pi^2 + 2i\pi\ln(2b)}{b}
    + \left( 2 + \frac{5\pi^2}{6} \right) 
    + {\cal O}(b) \,.
\end{aligned}
\end{equation}
Based on the symmetry properties of $F_1$ and $f_2$, it was concluded in \cite{Mitov:2009sv,Becher:2009kw} that these functions vanish whenever two of the velocities of the massive partons coincide. Indeed, this seems to be an obvious consequence of the fact that $F_1$ is totally anti-symmetric in its arguments, while $f_2$ is odd in its second argument. This reasoning implicitly assumes that the limit of equal velocities is non-singular, but is invalidated by the presence of Coulomb singularities in $r(\beta)$ and $g(\beta)$ near threshold. Consider the limit where two massive partons 1 and 2 are produced near threshold, with a small relative 3-velocity $\vec{v}_{12}\equiv\vec{v}_1-\vec{v}_2$ defined in their rest frame. It is then straightforward to derive that
\begin{equation}
   \lim_{v_2\to v_1} f_2
   = \frac{\alpha_s^2}{4\pi^2} 
   \left[ \pi^2 + 2i\pi\ln(2|\vec{v}_{12}|) \right] \cos\theta \,, 
\end{equation}
where $\theta$ is the scattering angle formed by the 3-momenta of particles 1 and 3. A similar formula holds for $F_1$. This result is anti-symmetric in the parton indices 1 and 2 as required, but it does not vanish at threshold. 

Another interesting limit is that of large recoil, where all the scalar products $w_{IJ}$ become large in magnitude. In that case, both $F_1$ and $f_2$ are suppressed like ${\cal O}(1/w^2)$, because for large $\beta$
\begin{equation}
   g(\beta)
   = \frac{1}{2w^2} \left[ \ln^2(2w) - \ln(2w) + \frac{\pi^2}{6} 
   - \frac12 \right] + {\cal O}\Big( \frac{1}{w^3} \Big) \,.
\end{equation}
Note that the non-planar contribution from the first graph in Figure~\ref{fig:dia}, which was studied in the Euclidean region in \cite{Mitov:2009sv}, contains the leading-power term
\begin{equation}
   F_1^{\rm non-planar} 
   = - \frac{\alpha_s^2}{12\pi^2}\,\ln\frac{w_{12}}{w_{23}}\,
    \ln\frac{w_{23}}{w_{31}}\,\ln\frac{w_{31}}{w_{12}} 
    + {\cal O}\Big(\frac{1}{w^2}\Big)
\end{equation}
and is unsuppressed in this limit. However, this contribution cancels against a leading-power term in the planar and counterterm contributions.

Using that $w_{IJ}=-s_{IJ}/(2m_I m_J)$, we see that the large-recoil limit corresponds to $m_I m_J\to 0$ at fixed $s_{IJ}$. It follows that the three-parton correlation terms described by $F_1$ and $f_2$ vanish like $(m_I m_J/s_{IJ})^2$ in the small-mass limit. This observation is in accordance with a factorization theorem proposed in \cite{Mitov:2006xs,Becher:2007cu}, which states that massive amplitudes in the small-mass limit can be obtained from massless ones by a simple rescaling prescription for the massive external legs.

\section{Anomalous dimension for $\bm{q\bar q\to t\bar t}$ near threshold}
\label{sec:tt}

As a sample application, we apply our formalism to the calculation of the two-loop anomalous-dimension matrices for top-quark pair production near threshold in the $q\bar q\to t\bar t$ channel. This matrix (along with the corresponding one in the $gg\to t\bar t$ channel) forms the basis for soft-gluon resummation at the next-to-next-to-leading logarithmic (NNLL) order. We adopt the $s$-channel singlet-octet basis, in which the $t\bar t$ pair is either in a color-singlet or color-octet state. For the quark-antiquark annihilation process $q_l(p_1)+\bar q_k(p_2)\to t_i(p_3)+\bar t_j(p_4)$, we thus choose the independent color structures as $c_1 = \delta_{ij}\,\delta_{kl}$ and $c_2 = (t^a)_{ij}\,(t^a)_{kl}$. In the threshold limit $s=2p_1\cdot p_2\to 4m_t^2$ it is convenient to define the quantity $\beta_t=\sqrt{1-4m_t^2/s}$, which is related to the relative 3-velocity $\vec{v}_{t\bar t}$ between the top-quark pair in the center-of-mass frame by $|\vec{v}_{t\bar t}|=2\beta_t$. We find that in the limit $\beta_t\to 0$ the two-loop anomalous-dimension matrices reduces to
\begin{equation}\label{Gqqthresh}
\begin{split}
   \bm{\Gamma}_{q\bar q} 
   &= \bigg[ C_F\,\gamma_{\rm cusp}(\alpha_s)
    \left( \ln\frac{s}{\mu^2} - \frac{i\pi}{2\beta_t} - i\pi + 1
    \right) \\
   &\quad\mbox{}+ C_F\,\gamma_{\rm cusp}^{(2)}(\beta_t)
    + 2\gamma^q(\alpha_s) + 2\gamma^Q(\alpha_s) \bigg]
    \begin{pmatrix}
     1~ & ~0 \\ 0~ & ~1
    \end{pmatrix} 
    \nonumber 
\end{split}
\end{equation}
\begin{equation}
\begin{split}    
   &\mbox{}+ \frac{N}{2} \left[ \gamma_{\rm cusp}(\alpha_s)
    \bigg( \frac{i\pi}{2\beta_t} + i\pi - 1 \bigg)
    - \gamma_{\rm cusp}^{(2)}(\beta_t) \right]\!
    \begin{pmatrix}
     0~ & ~0 \\ 0~ & ~1
    \end{pmatrix} \\
   &\mbox{}+ \frac{\alpha_s^2}{2\pi^2}
    \left[ \pi^2 + 2i\pi\ln(4\beta_t) \right] \cos\theta
    \begin{pmatrix}
     0 & \frac{C_F}{2} \\ -N & 0
    \end{pmatrix}
    + {\cal O}(\beta_t) \,,
\end{split}
\end{equation}
where the two-loop expressions for the anomalous dimensions $\gamma_{\rm cusp}$, $\gamma^q$, and $\gamma^Q$ can be found in \cite{Becher:2009kw}, and 
\begin{equation}
   \gamma_{\rm cusp}^{(2)}(\beta_t) 
   = \frac{N\alpha_s^2}{2\pi^2} 
    \left[ \frac{i\pi}{2\beta_t} \left( 2 - \frac{\pi^2}{6} \right) 
    - 1 + \zeta_3 \right]  
\end{equation}
arises from the threshold expansion of the two-loop coefficient of the velocity-dependent cusp anomalous dimension $\gamma_{\rm cusp}(\beta,\alpha_s)$.

We stress that, as a consequence of the Coulomb singularities, the three-parton correlation term does not vanish near threshold. Instead, it gives rise to scattering-angle dependent, off-diagonal contribution in (\ref{Gqqthresh}). The off-diagonal terms were omitted in two recent papers \cite{Beneke:2009rj,Czakon:2009zw}, where threshold resummation for top-quark pair production was studied at NNLL order. We leave it to future work to explore if and how the results obtained by these authors need to be modified in light of our findings.

\section{Conclusions}
\label{sec:concl}

The IR divergences of scattering amplitudes in non-abelian gauge theories can be absorbed into a multiplicative renormalization factor, whose form is determined by an anomalous-dimension matrix in color space. At two-loop order this anomalous-dimension matrix contains pieces related to color and momentum correlations between three partons, as long as at least two of them are massive. This information is encoded in two universal functions: $F_1$, describing correlations between three massive partons, and $f_2$, describing correlations between two massive and one massless parton. In this Letter we have calculated these functions at two-loop order. Using the exact analytic expressions, we studied the properties of the three-parton correlations in the small-mass and threshold limits. We found that the functions $F_1$ and $f_2$ vanish as $(m_I m_J/s_{IJ})^2$ in the small-mass limit, in accordance with existing factorization theoreoms for massive scattering amplitudes \cite{Mitov:2006xs,Becher:2007cu}. On the other hand, and contrary to naive expectations, the two functions do not vanish in the threshold limit, because Coulomb singularities compensate a zero resulting from the anti-symmetry under exchange of two velocity vectors. This fact has been overlooked in the recent papers \cite{Beneke:2009rj,Czakon:2009zw}, where the three-parton correlations were neglected near threshold.

Our results allow for the calculation of the IR poles in an arbitrary on-shell, $n$-particle scattering amplitude to two-loop order, where any number of the $n$ partons can be massive. As an application, we have derived the anomalous-dimension matrix for top-quark pair production in the $q\bar q\to t\bar t$ channel. We will explore in future work to what extent the new off-diagonal entries, arising from three-parton correlation terms, affect the numerical results for the threshold-resummed $t\bar t$ production cross sections at the Tevatron and LHC. 

This Letter completes the study of IR divergences of two-loop scattering amplitudes with an arbitrary number of massive and massless external particles, and in arbitrary non-abelian (or abelian) gauge theories with massless gauge bosons. Details of our calculations will be presented in a forthcoming article. 

{\em Acknowledgments:\/}
We are grateful to Thomas Becher for collaboration during the early stages of this work.

\end{document}